# Ambipolar Surface State Thermoelectric Power of Topological Insulator $Bi_2Se_3$

*Dohun Kim[1†], Paul Syers[1], Nicholas P. Butch[2], Johnpierre Paglione[1], and Michael S. Fuhrer[1,3*]*

1. Center for Nanophysics and Advanced Materials, Department of Physics, University of Maryland, College Park, MD 20742-4111, U.S.A.

2. Center for Neutron research, National Institute of Standards and Technology, Gaithersburg, MD 20899-6102, U.S.A.

3. School of Physics, Monash University, 3800 Victoria, Australia

ABSTRACT

We measure gate-tuned thermoelectric power of mechanically exfoliated $Bi_2Se_3$ thin films in the topological insulator regime. The sign of the thermoelectric power changes across the charge neutrality point as the majority carrier type switches from electron to hole, consistent with the ambipolar electric field effect observed in conductivity and Hall effect measurements. Near charge neutrality point and at low temperatures, the gate dependent thermoelectric power follows the semiclassical Mott relation using the expected surface state density of states, but is larger than expected at high electron doping, possibly reflecting a large density of states in the bulk gap. The thermoelectric power factor shows significant enhancement near the electron-hole puddle carrier density ~ 0.5 x $10^{12}$ $cm^{-2}$ per surface at all temperatures. Together with the expected reduction of lattice thermal conductivity in low dimensional structures, the results demonstrate that nanostructuring and Fermi level tuning of three dimensional topological insulators can be promising routes to realize efficient thermoelectric devices.

KEYWORDS  Topological Insulators, Thermoelectric Power, Ambipolar Electric Field Effect, Bismuth Selenide, Charge Transfer Doping.

The efficiency of thermoelectric devices is determined by the thermoelectric figure of merit $ZT = \sigma S^2 T/\kappa$, where $S$ is thermoelectric power (Seebeck coefficient) and $\sigma$ and $\kappa$ are electrical and thermal conductivity respectively. For a given thermal conductivity (typically dominated by phonons) and temperature, $ZT$ is determined by the electronic structure of a material through the power factor $\sigma S^2$. Nanostructuring has long been seen as a promising route to increase $ZT$ both through decreasing $\kappa$ via phonon boundary scattering, and increasing the power factor through confinement-induced band structure effects[1]. The discovery that important thermoelectric materials $Bi_{1-x}Sb_x$ and $Bi_2(Se,Te)_3$ [2-4] are in fact 3D topological insulators (TI) has triggered new directions to improve $ZT$ via nanostructuring. The 3D TIs have an insulating bulk but metallic surface states with spin-momentum helicity which cannot be localized by non-magnetic disorder[4,5]. Recent theoretical calculations[6,7] show that significant enhancement of $ZT$ is expected in 3D TIs particularly in thin film geometry where top and bottom surfaces can hybridize and form a surface gap[8].

Thermoelectric power is also of great interest in understanding the electronic transport properties of Dirac electronic systems[9], thus measurement of the thermoelectric properties of TI surface states can elucidate details of the electronic structure of the ambipolar nature of topological metallic surface states that cannot be probed by conductance measurements alone[10]. Although there have been theoretical investigations of thermal and thermoelectric properties of Dirac materials including graphene and TIs[6,7,11-15] and thermoelectric transport in carbon nanotubes[16] and graphene[9,17-19] have been studied experimentally, it has been difficult to experimentally observe clear surface thermoelectric transport in TI materials due to the high level of bulk doping present in as-grown TI crystals[20,21].

Here we experimentally demonstrate the first measurement of thermoelectric power of topological insulator $Bi_2Se_3$ dominated by metallic surface states. We show a clear ambipolar electric field effect indicated by the sign change of the thermoelectric power across the charge neutrality point where the charge transport exhibits minimum conductivity. We find that the gate dependent thermoelectric power exhibits an agreement with the semiclassical Mott relation[10] consistent with surface band structure, however, we also find deviation with expected surface dominated thermoelectric power at high electron doping voltage where bulk states are likely populated. We also demonstrate significant (up to two-fold) enhancement of the power factor $\sigma S^2$ for a Dirac band as the Fermi energy approaches the Dirac point, indicating that surface state conduction in TI materials is a promising route to thermopower enhancement, even in the absence of a hybridization gap[6,7]. We also discuss possible routes for further enhancement to the limit predicted theoretically[6,7].

We study mechanically exfoliated $Bi_2Se_3$ single crystals[20] with thickness approximately $13 \pm 1$ *nm* which is confirmed using atomic force microscopy (AFM). We use a thermopower measurement technique using micro-fabricated heaters and thermistors, previously developed and used to study thermoelectric properties of carbon nanotubes[16] and graphene[9,18,19] (see also Methods for description of measurement scheme). As described previously[22], we employed molecular charge transfer *p*-type doping by thermal evaporation of 2,3,5,6-tetrafluoro-7,7,8,8-tetracyanoquinodimethane to reduce high level of *n*-type doping. Further tuning of carrier density is accomplished using a Si back gate and 300 *nm* SiO$_2$ gate dielectric. We showed previously [22] that in such thin Bi$_2$Se$_3$ slabs in the TI regime that the top and bottom surfaces are strongly capacitively coupled through the high-dielectric-constant bulk ($\varepsilon$ ~100 [20]) and couple nearly equally to the back gate.

Figures 1a and b show sheet conductivity σ and thermopower *S* respectively as a function of back gate voltage $V_g$ for a representative Bi$_2$Se$_3$ device for several temperatures *T* from 10 to 240*K*. We measure Hall carrier density $n_H$ about 3 x 10$^{12}$ *cm*$^{-2}$ at zero gate voltage (see Fig. 1a, inset) at the temperature of 10 *K* with the highest Hall mobility at low temperature of ~1100 cm$^2$/Vs. The slope $n_H$ vs. $V_g$ reflects the expected capacitance of ~11 nFcm$^{-2}$ for the 300 nm SiO$_2$ gate dielectric. At *T* = 10 *K* the device shows clear ambipolar electric field effect which is indicated by minimum conductivity $\sigma_{min}$ at charge neutrality point $V_D$ ~ -50 V and corresponding sign change of Hall carrier density $n_H$ (see Fig. 1a, inset) consistent with gapless Dirac band structure of topological surface states[22]. As it was shown previously[22-24], σ is linearly proportional to $V_g$ near $V_D$ suggesting that charged impurity scattering limits electronic transport for the present devices[22]. The observation of sublinear σ($V_g$) at $V_g$ further away from $V_D$ may indicate that there are additional types of disorder, e.g. neutral point defects, which need to be considered[25], though as discussed further below it may also reflect the population of lower-conductivity bulk states, most likely by filling in-gap impurity bands. At low temperatures T < 150 *K*, we observe a change in sign of *S*, which indicates the sign of the majority charge carrier from positive to negative as $V_g$ crosses $V_D$, consistent with the change in sign of $n_H$. At temperatures starting from about 70 *K*, thermal population of hole and electron carriers between the bulk valence band and the surface state bands becomes important[26], resulting in a shift of $V_D$ to negative voltage, reflected also in a shift to negative voltage of the zeroes in *S*. This thermal activation effect was explained by us in detail in a previous work [26]. At higher temperatures T ≥ 150 *K* thermal activation is significant enough that the sign change cannot be observed within the range of accessible $V_g$. The electron-hole asymmetry observed in both $\sigma$ and *S* is consistent with the asymmetric surface band structure of $Bi_2Se_3$ [3,27]. The peak value of *S* reaches ~170 $\mu V/K$ at 240 *K* in the majority of devices measured.

We now turn to a quantitative analysis of gate-tuned surface state thermopower in the TI regime. For semiclassical transport the carrier-density dependence of the thermopower is described by the Mott relation[10]

$$S = -\frac{\pi^2 k_B^2 T}{3e} \frac{d \ln \sigma}{dE_F} = -\frac{\pi^2 k_B^2 T}{3e} \frac{1}{\sigma} \frac{d\sigma}{dV_g} \frac{dV_g}{dE_F} = -\frac{\pi^2 k_B^2 T}{3e} \frac{1}{\sigma} \frac{d\sigma}{dn} D(V_g) \ , \qquad (1)$$

where *e* is the elementary charge, $n = C_g(V_g - V_D)/e$ is gate induced carrier density with gate capacitance $C_g \approx 11.1$ nFcm$^{-2}$, $k_B$ is the Boltzmann constant, $E_F$ is Fermi energy, and $D(V_g)$ is the gate-voltage-dependent density of states. The first equality can be tested with concurrently measured $\sigma(V_g)$ along with expected band structure and has been shown to provide a good description of $S(V_g)$ in mesoscopic carbon nanotube[16] and graphene[9,17,18] field-effect devices. For a Dirac material with Coulomb scattering, Eqn. (1) predicts a divergence in $S \sim (V_g - V_D)^{-1/2}$, since $\sigma \sim (V_g - V_D)$ and $E_F \sim (V_g - V_D)^{1/2}$ [9,15]. This enhancement is the essence of the enhanced thermopower expected for Dirac surface states. The divergence is cut off by the electron-hole puddle carrier density *n\**; we can estimate $n^* \sim 0.5 \times 10^{12}$ cm$^{-2}$ per surface from the minimum in $n_H$ (see Fig. 1a, inset).

For a detailed comparison with Eqn. (1), we first compare the measured $S$ with predicted $S$ ($S_{Mott}$) using known surface band structure of $Bi_2Se_3$ [3, 27]. We calculate the factor $dV_g/dE_F$ numerically using a band model as developed previously[26] using the surface band with dispersion $E_{2D}(k) = \hbar v_{F,0} k + \frac{\hbar^2 k^2}{2m_s^*}$. For the calculation, we used the parameters of the Fermi velocity near Dirac point $v_{F,0} = 3 \times 10^7$ cm/s, and effective mass $m_s^* = 0.3\, m_e$ of the surface band which are reasonable for $Bi_2Se_3$ from the ARPES measurements[3,27]. We also calculated expected carrier density using surface band and bulk valence and conduction bands. For the bulk band density of states we used first principles calculation of bulk $Bi_2Se_3$ reported elsewhere[28].

Figure 2a shows a detailed comparison of the measured thermopower ($S$, black solid) and calculated surface state thermopower from semi-classical Mott relation ($S_{Mott}$, red solid) using conductivity data and band model described above for various temperatures from 10 to 240 $K$. We find a good agreement at low temperature (e.g. 10 $K$) and near charge neutrality point, as expected for degenerate electrons in the metallic surface band where the entropy transported in the channel is proportional to the number of thermally activated carriers over the degenerate Fermi sea ($S \sim T/E_F$) [9]. This agreement is also true near $V_D$ due to charge inhomogeneity driven puddle density $> 10^{12}$ cm$^{-2}$ which renders the system to remain in the degenerate regime. We observe a peak in $S$ at a gate voltage roughly corresponding to the electron-hole puddle carrier density $n^*$, below which the value of $S$ decreases due to cancellation of $n$-type and $p$-type thermopowers, with a magnitude which can be explained by the Mott relation [Eqn. (1)], as observed previously for graphene[9]. The Mott relation also reproduces the position and magnitude of some of the fine features in the low-temperature (10 and 20K) gate-voltage dependent thermopower (presumably due to universal conductance fluctuations). Figure 2b shows calculated surface (black curves) and bulk (blue curves) carrier densities at given total charge density fixed by $V_g$ as a function of $V_g$ at the same temperatures in Fig 2a. We observe significant thermal activation at the temperatures T ≥ 150 *K* near or below -50*V* (Fig. 2b shaded area) as explored in detail previousely[26]. However, for positive $V_g$ up to $V_g$=80*V*, thermal activation to bulk conduction band is negligible at all temperatures.

We also observe an additional enhancement of the thermopower which is not predicted by the Mott relation assuming only surface states. This can be seen especially at high positive $V_g$ and high temperature where $S_{Mott}$ tends to zero but the measured *S* saturates to a constant. The ratio $S/S_{Mott}$ is as large as 7 at *T* = 240 K and $V_g$ = +80 V. Figures 3a and b show the temperature dependence of the measured *S* and predicted $S_{Mott}$ at fixed $V_g$ ranging -80 < $V_g$ < 80*V* indicated in the caption, and Figure 3c shows the additional thermopower $S - S_{Mott}$. The additional thermopower is negative, with magnitude increasing monotonically with temperature, and roughly independent of gate voltage. Phonon drag is a known contribution to thermopower beyond Eqn. (1). However, the phonon drag contribution to the thermopower $S_p$ in a Dirac electronic system was found to obey Herring's law $S_p = -v_s \Lambda/\mu_p T$ where $v_s$ is the sound velocity, $\Lambda$ the phonon mean free path, and $\mu_p$ is the phonon-limited carrier mobility[29]. We expect at high temperature $\Lambda \sim 1/T$, $\mu_p \sim 1/nT$ [26], and thus $S_p \sim n/T$, in sharp contrast to the experimental observation of additional thermopower independent of *n* and increasing with *T* (roughly proportional to $T^2$ at high *T*). Similarly, temperature-dependent screening gives corrections to the conductivity and thermopower proportional to $(k_B T/E_F)^2$ [15]; such corrections should be unimportant at high $V_g$ [26], and moreover no strong dependence on $E_F$ or $V_g$ is seen. By measuring an additional sample with factor of three smaller contact area we found quantitatively similar temperature-dependent thermopower (within 5µV/K), indicating that the effect of contact barrier on the additional contribution to the measured thermopower is small (see Supplementary Information). In the Supplementary Information, we also show the negligible temperature dependence of the carrier density in the electron-doped regime for a similarly prepared sample, which precludes temperature dependent charge transfer as the source of additional contribution to the thermopower.

We expect that the Mott formula is a good approximation for temperatures $k_BT < E_F$. Due to electron-hole puddling the effective Fermi energy is never significantly smaller than 50 meV, and we expect this approximation to be reasonable for the temperature range studied. Furthermore the approximation holds for multiband system, and the Mott relation (second equality in Eqn. [1]) can be inverted using the measured $S$ and $\sigma(V_g)$ to obtain a total density of states $D$. Figure 4 shows a comparison of the experimentally obtained gate-dependent density of states from inversion of the Mott formula at various temperatures (solid lines) with the expected surface[3,27] (dashed lines) and bulk[28] (dash-dot lines) densities of states with bulk valence band edge $E_v \approx$ -60meV respect to the Dirac point [3, 26, 27]. Near charge neutrality point of ≈-50V and at low temperatures, the measured $D$ agrees well with expected surface density of states for $Bi_2Se_3$ as expected, reflecting the agreement seen in Fig. 2. Away from charge neutrality point, a large enhancement of $D$ is seen in the conduction band at low temperatures, well below the expected onset of the bulk conduction band density of states at $V_g$ >80$V$. This possibly reflects a large smearing of the conduction band edge, or the presence of a high density of impurity states in the bulk gap that are not reflected in the carrier density calculation in Fig. 2b. We note that the extraction of $D$ using the Mott relation does not determine whether the states are conducting or insulating. However, Hall carrier density measurements suggest that the vast majority of the gate-induced charge is delocalized. If bulk states contribute to the conduction at large electron doping, this could provide an alternative explanation for the sublinearity of $\sigma(V_g)$ observed in the same regime.

On raising the temperature, we observe a filling-in of the dip in $D$ associated with the Dirac point. This is qualitatively consistent with thermal smearing. The larger $D$ implies that our gate voltage window accesses a smaller Fermi energy range, therefore larger thermal activation effects are expected. This would also explain the unexpected super-linear $S(T)$ seen in Fig. 3. While the presence of an unexpectedly large $D$ inside the bulk gap qualitatively explains the observed features of the data, a quantitative understanding will require a more detailed knowledge of the origin of such states. Particularly if the states are inhomogeneously distributed through the sample, due to screening of disorder or "puddling", or due to band-bending, then the extraction of $D$ is only qualitative. We conclude that the thermopower very close to the Dirac point and at low temperature is explained well by the expected $D$ and measured $\sigma(V_g)$ of the surface states alone. Away from the Dirac point, and at higher temperature, deviations are observed which are presumed to be due to additional states in the bulk band gap along with increased thermal activation to these states.

We finally turn to discuss enhanced thermoelectric efficiency in the TI regime. A simple analysis for the Dirac surface state predicts $S \sim (V_g - V_D)^{-1/2}$, $\sigma \sim (V_g - V_D)$ hence power factor $\sigma S^2$ is independent of $V_g$. In recent theoretical studies[6,7], a large enhancement of the power factor is predicted at low Fermi energies when considering thin slabs of $Bi_2Se_3$ and $Bi_2Te_3$ in which inter-surface hybridization induces a surface gap, but no enhancement of the power factor is predicted for ungapped surface states near the Dirac point[6]. Figure 5 shows the measured $\sigma S^2$ as a function of $V_g$ at temperatures from 20 to 240 $K$. Notably, a clear enhancement of $\sigma S^2$ can be observed near $V_g = -40V$ at all temperatures, which corresponds to carrier density $n \sim 1 \times 10^{12}$ cm$^{-2}$ or $E_F \sim 40$ meV with respect to the Dirac point assuming a linear Dirac band for Bi$_2$Se$_3$. At all temperatures the enhancement of $\sigma S^2$ is about factor of two compared to the non-TI regime (bulk conducting). The enhancement near the Dirac point is surprising given that we expect the hybridization gap in our 13 nm thick samples to be negligible (~ 10 μeV [8]). The maximum $\sigma S^2$ of ~ 3 μWcm$^{-1}$K$^{-2}$ measured at 240 K is an order of magnitude enhancement compared to the value of ~ 0.3 μWcm$^{-1}$K$^{-2}$ at 300 K reported for Bi$_2$Se$_3$ nanoflakes (highly conducting bulk) prepared by solvothermal method[30]. Admittedly, the absolute value of maximum $\sigma S^2$ for the present samples are less then best reported value ~ 18 μWcm$^{-1}$K$^{-2}$ of bulk Bi$_2$Se$_3$ [31] at the same temperature.

By controlling doping of exfoliated Bi$_2$Se$_3$ we experimentally confirmed the gapless ambipolar nature of the thermopower that is consistent with the existence of topological metallic surface states. Although more works are needed to quantitatively understand the observed deviation of measured $S$ from the semicalssical Mott relation, we find a clear enhancement of $\sigma S^2$ near the Dirac point at all temperatures. Since one additionally expects a reduction in thermal conductivity for thin Bi$_2$Se$_3$ compared to bulk[32], we find that nanostructuring and Fermi level control of topological insulators are promising strategies to increase $ZT$ by both increasing the power factor and reducing the thermal conductivity. While the power factor for the Bi$_2$Se$_3$ surface states does not exceed the bulk in our samples, there is significant room for improvement; we expect that the enhancement of power factor in thin topological insulators in the topological regime can be improved by increasing surface quality (increase electrical conductivity as well as $S$ by reducing $n^*$ and thus more closely approaching the Dirac point). For current samples having thickness ≈13 nm we estimate the hybridization gap is in the order of ≈10 μeV [8], much smaller than charge inhomogeneity driven energy broadening, which limits the thermopower enhancement by a factor of two. Further thinning of the samples to open a sizeable hybridization gap between top and bottom surfaces[6,8] as well as further reducing the lattice thermal conductivity[32] could produce a larger enhancement of thermoelectric performance.

**Methods**

**$Bi_2Se_3$ thermopower device fabrication** Low-doped (carrier density ~ $10^{17}$ $cm^{-3}$) bulk $Bi_2Se_3$ single crystals with bulk resistivity exceeding 2 *mΩcm* at 300 *K* were grown by melting high purity bismuth (6*N*) and selenium (5*N*) in sealed quartz ampoules[20]. The crystals were exfoliated with Scotch tape and deposited on doped Si covered with 300 *nm* $SiO_2$. Thin $Bi_2Se_3$ crystals with thickness in the order of ~10 *nm* were identified by combined use of optical and atomic force microscopy (AFM). Electron beam lithography, thermal evaporation and liftoff techniques were used to make electrical contact, thermistors and external heaters (Cr/Au: 5/70 nm). Typical line widths of thermistors were restricted to be less than 500 *nm* to ensure local thermometry. Thin films were patterned into Hall bar geometry (see Fig. 1b, inset) with typical aspect ratio (*L*/*W*) of about 2 and shortest length exceeding 2 *μm* using Ar plasma at a pressure of ~6.7 Pa (5 x $10^{-2}$ Torr). Molecular charge transfer doping was done by thermal evaporation of ~10 *nm* of 2,3,5,6-tetrafluoro-7,7,8,8-tetracyanoquinodimethane organic molecules (SynQuest Labs) on top of the fabricated samples[22,33].

**Measurement** Thermopower measurement was done using Stanford Research Systems SR830 Lock in amplifiers and a commercial cryostat equipped with 9 *T* superconducting magnet. Low frequency ac (*ω* < 17 Hz) heater currents was applied to the sample and the resultant 2*ω* thermoelectric voltage Δ*V* between two ends of thermisters (see Fig. 1b, inset) was detected in the open circuit condition[9,16]. Temperature difference $\Delta T$ was measured by measuring temperature-calibrated four probe resistances of the two gold thermistors (see Fig. 1b, inset). Finally thermoelectric power was calculated from $S = -\Delta V/\Delta T$. Thermopower of Au leads (~2 $\mu V/K$ at 300 $K$) at a given temperature is expected to be negligible compared to that of $Bi_2Se_3$ and was not subtracted from the measured *S*. Carrier density was tuned through 300 *nm* thick $SiO_2$ dielectric back gate. To stay in the linear response regime, heater current was adjusted (typically 3 to 8 *mA*) so that the condition $\Delta T << T$ at given temperature is satisfied[9]. Electronic transport measurements were performed using similar four probe lock-in technique with rms current amplitude of 100 *nA* applied to the device and measuring first harmonic ($\omega$) voltage signals.

**Contents figure**

Ambipolar Surface State Thermoelectric Power of Topological Insulator Bi$_2$Se$_3$

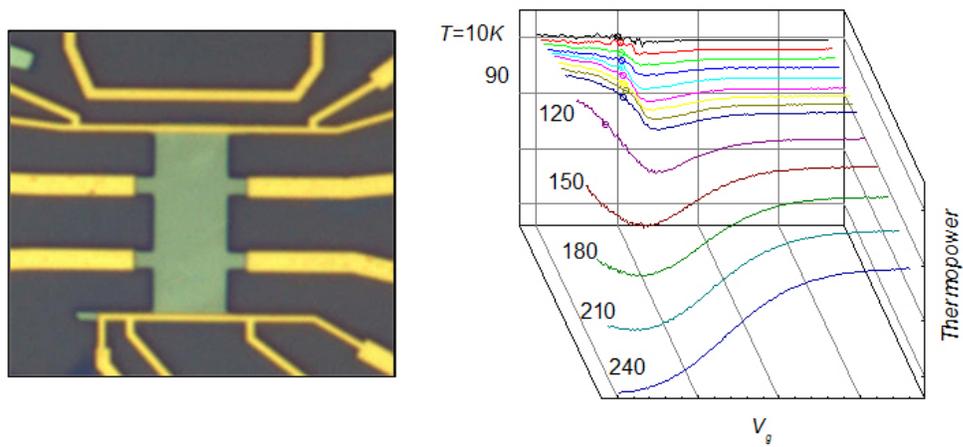

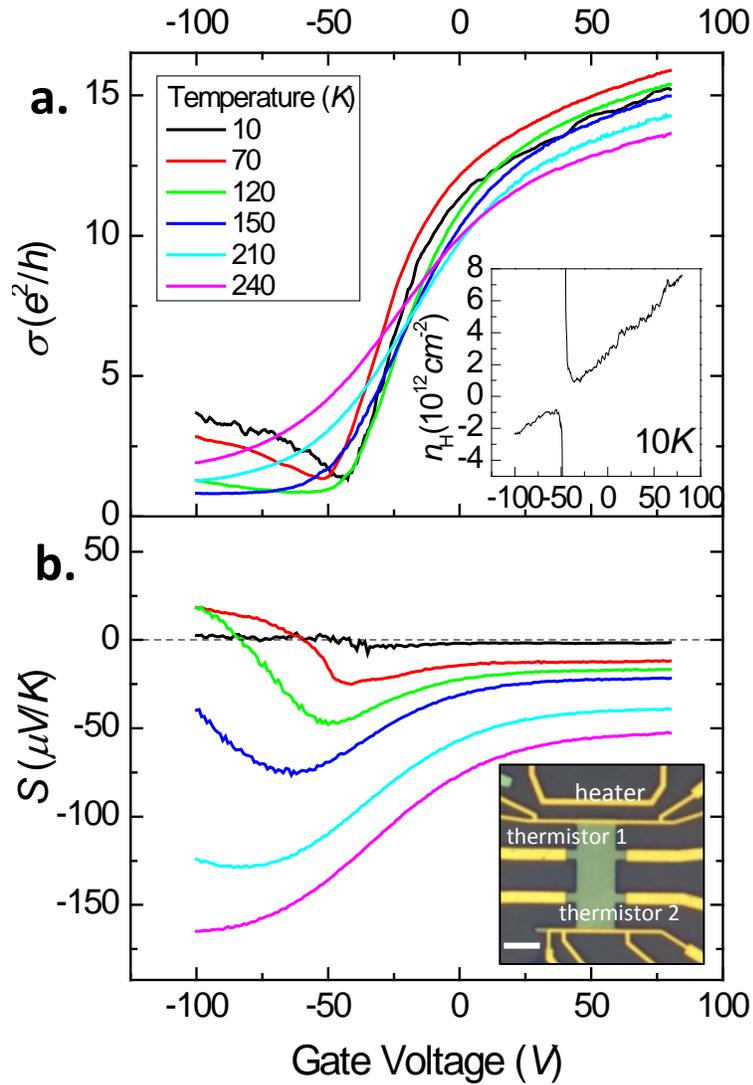

**Figure 1. a,** 2D conductivity $\sigma$ as a function of back gate voltage $V_g$ of a ~13 *nm* thick $Bi_2Se_3$ measured at temperatures indicated in the caption. The inset shows Hall carrier density $n_H$ simultaneously measured with $\sigma$ at the temperature of 10 *K*. **b,** Thermopower $S$ as a function of $V_g$ measured at the same condition as in **a** showing ambipolar electric field effect consistent with corresponding $\sigma$ measurement. The inset shows optical micrograph of the device. The scale bar is 2 *μm*.

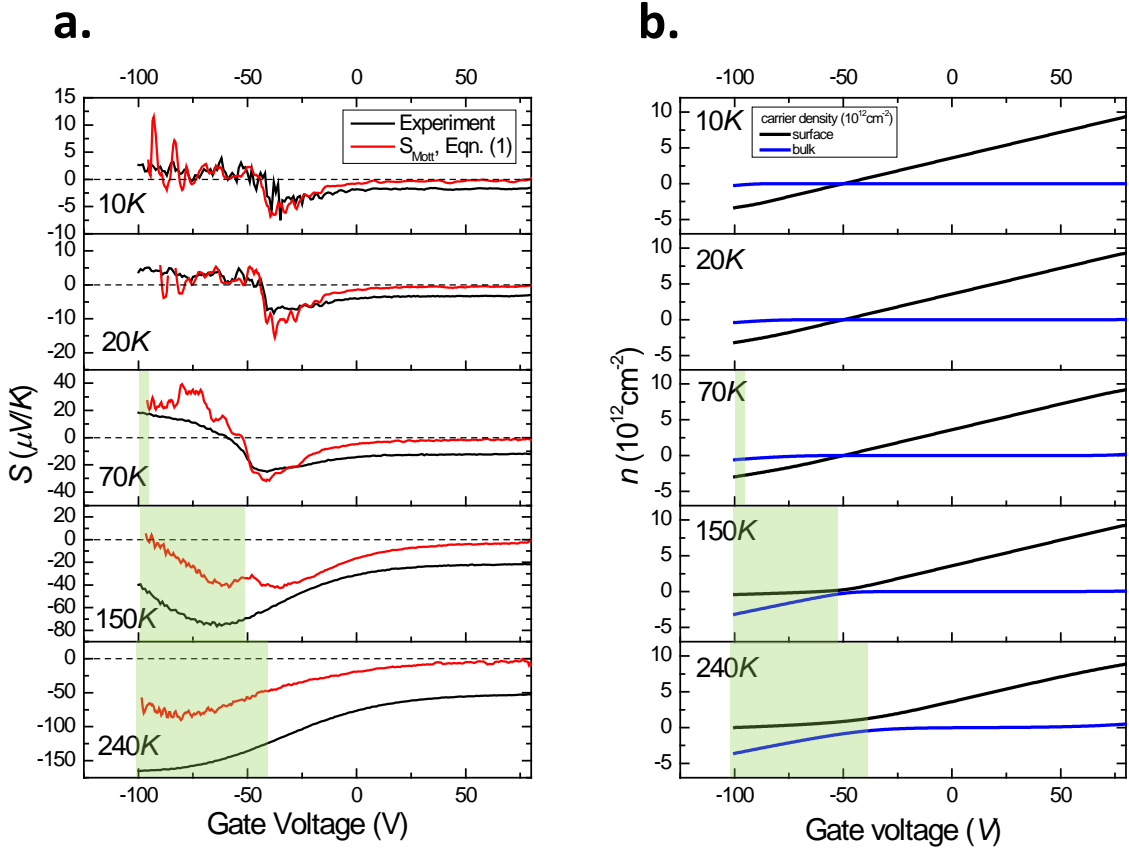

**Figure 2. a.** Experimentally measured thermopower $S$ (black solid curve) and thermopower estimated from the Mott relation $S_{Mott}$ including effect of thermal activation of carriers from bulk valence band and shift of Dirac point (see main text for discussion) as a function of $V_g$ at the temperatures from 10 to 240 $K$. Dashed horizontal lines denote zeros of thermopower. Shaded areas show gate voltage regions where thermal activation of bulk carriers is significant. **b.** Calculated surface (black curves) and bulk (blue curves) carrier densities as a function of $V_g$ at the same temperatures as in **a.** using band model described in the main text.

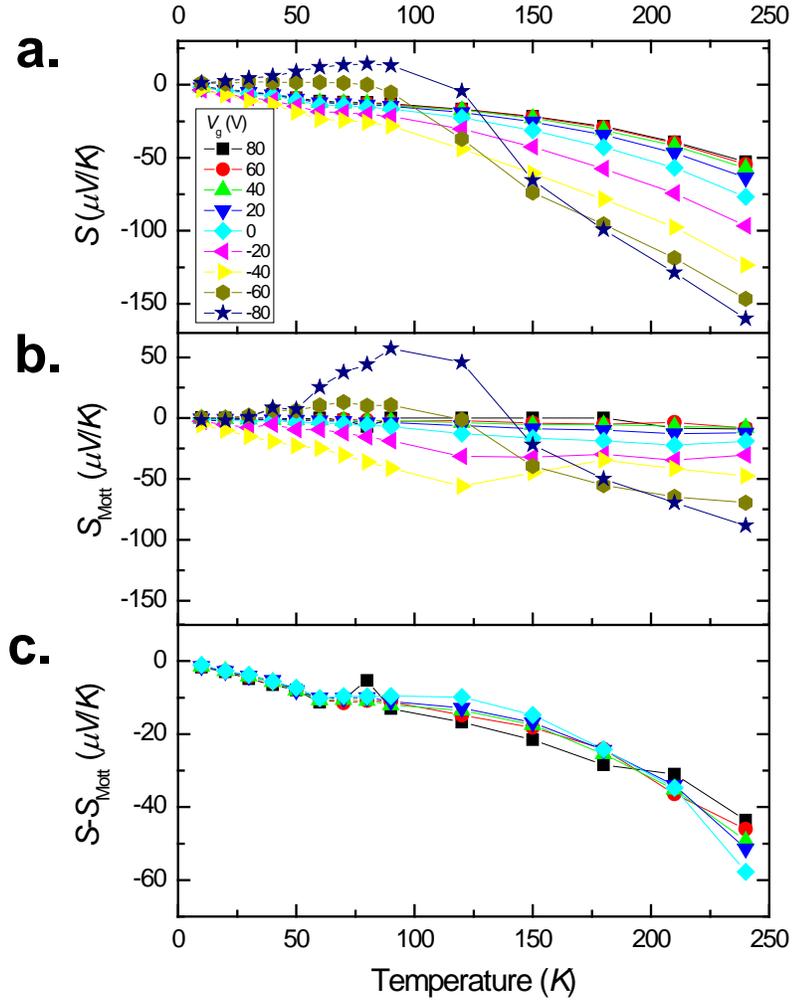

**Figure 3.** Variation of the **a,** measured thermo-power $S$ and **b,** predicted thermopower $S_{\text{Mott}}$ at fixed $V_g$ ranging $-80 < V_g < 80V$ indicated in the caption as a function of temperature from 10 to 240 $K$ showing nonlinear and non-monotonic temperature dependence. **c,** $S$ - $S_{\text{Mott}}$ as a function of temperature at the same $V_g$ given in **a** and **b**.

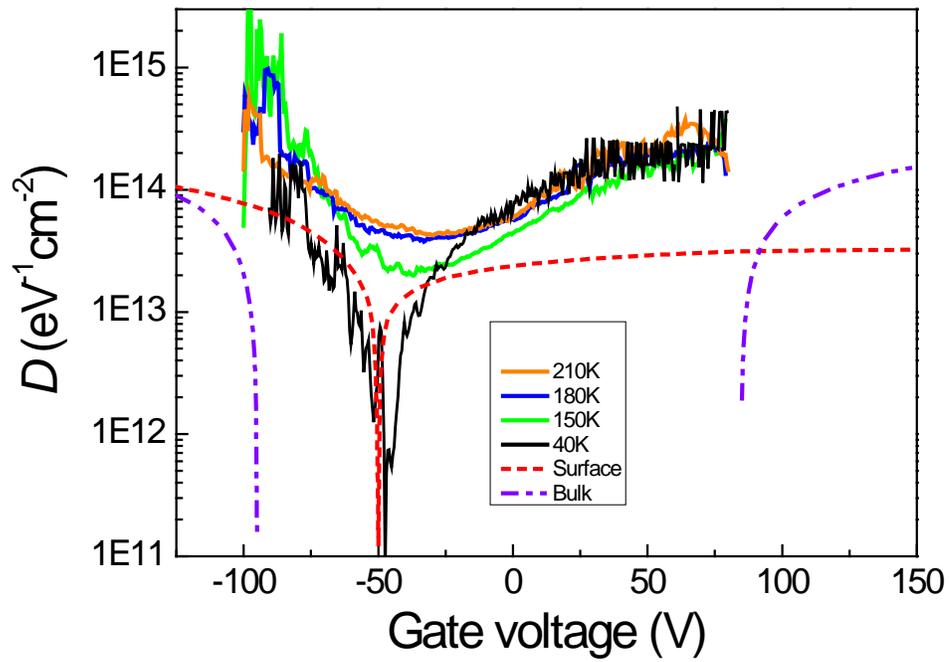

**Figure 4.** Density of states $D$ estimated from Mott relation (Eqn. 1) as a function of gate voltage $V_g$ at temperatures indicated in the caption. Dashed red (dash-dot purple) curves show expected surface (bulk) density of states of $Bi_2Se_3$.

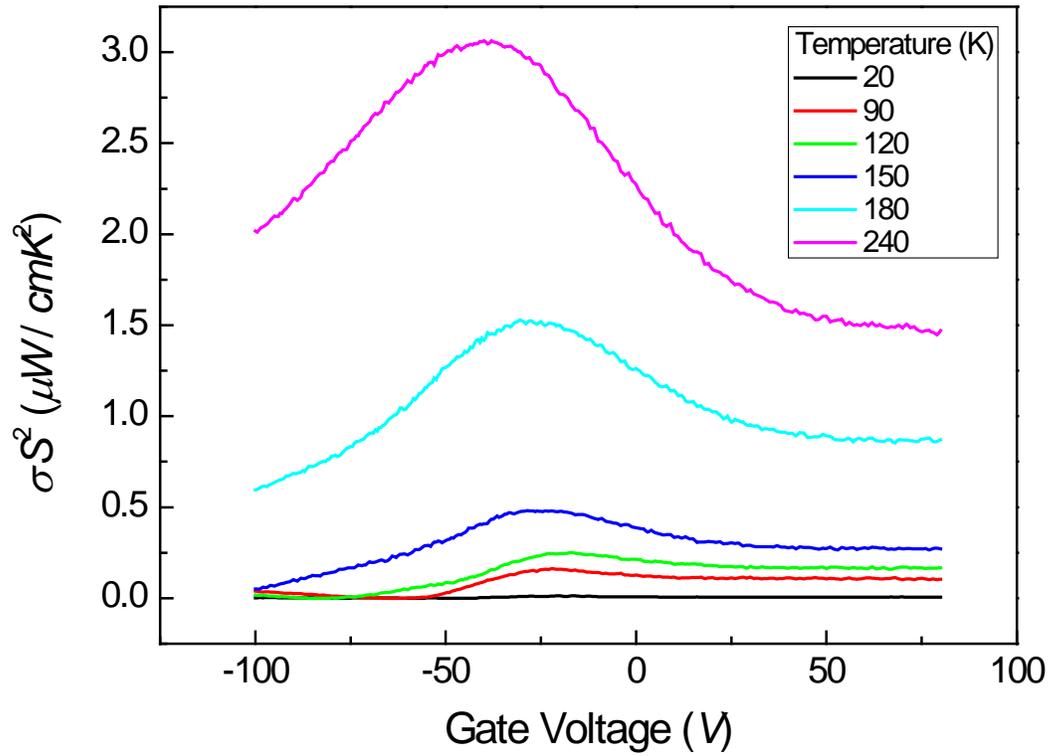

**Figure 5.** Thermoelectric power factor $\sigma S^2$ as a function of $V_g$ at various temperatures indicated in the caption. A factor of two enhancement of $\sigma S^2$ is observed near $V_g \sim -40V$ corresponding to total carrier density of $\sim 1 \times 10^{12}\, cm^{-2}$

**AUTHOR INFORMATION**

**Corresponding Author**     *E-mail: michael.fuhrer@monash.edu

**Note** The authors declare no competing financial interests.

**Present address**

† Department of Physics, University of Wisconsin-Madison, Madison, Wisconsin 53706, USA

**Author Contributions**

D.K. fabricated devices, performed the electrical measurements, and analyzed the data with M.S.F. P.S., N.P.B, and J.P. prepared single crystal $Bi_2Se_3$ starting material. D.K. and M.S.F. wrote the manuscript with contributions from all authors.

**ACKNOWLEDGMENT**

This work was supported by NSF grant number DMR-1105224. Preparation of $Bi_2Se_3$ was supported by NSF MRSEC (DMR-0520471) and DARPA-MTO award (N66001-09-c-2067). NPB was partially supported by the Center for Nanophysics and Advanced Materials. MSF is supported by an ARC Laureate Fellowship.

**Supporting Information**

Thermoelectric measurements on additional sample, reproducibility of additional contribution to thermopower, and F4TCNQ charge doping stability. This material is available free of charge via the Internet at http://pubs.acs.org.